# Brownian motion of molecules: the classical theory

Roumen Tsekov
Department of Physical Chemistry, University of Sofia, 1164 Sofia, Bulgaria

A short review of the classical theory of Brownian motion is presented. A new method is proposed for derivation of the Fokker-Planck equations, describing the probability density evolution, from stochastic differential equations. It is also proven via the central limit theorem that the white noise is only Gaussian.

Methods of the non-equilibrium statistical mechanics are main tools for solving many theoretical problems of the contemporary physical chemistry. Their aim is to obtain the spatial and temporal evolution of the macroscopic properties of matter on the basis of the mechanical laws governing the molecular motion. Historically, this program has been realized in two directions: the kinetic theory of dilute gases and the theory of Brownian motion. Both of them have played major role in the development of the statistical mechanics and represent the origin of all modern statistical methods. The main achievement of the kinetic theory of dilute gases is the famous Boltzmann equation, which even nowadays is a primary key for understanding of the behavior of many-particles systems [1, 2]. On its basis one can express all transport coefficients via molecular characteristics. However, it is a very complex mathematical problem to solve the Boltzmann nonlinear integro-differential equation even in the case of extremely simple model systems. In contrast to the kinetic theory of dilute gases the theory of Brownian motion operates with easier mathematical apparatus and provides elegant solutions of many relaxation problems. In fact, it is the first attempt for stochastic modeling of the Nature, which determines it's very important contribution to the theory of random processes in physics, chemistry and biology [3-5] as well as in mathematics [6, 7]. On the basis of this theory powerful methods for solving various relaxation problems have arisen such as the theories of Markov stochastic processes [8, 9], of stochastic differential equations, of diffusive random processes [10, 11], etc. [12-14]. However, the theory of Brownian motion is restricted by two main limitations. First, it is a phenomenological theory in the frames of which it is not possible to express the involved kinetic coefficients via molecular characteristics. Second, the theory of Brownian motion postulates a tendency of mechanical systems towards thermodynamic equilibrium with a known equilibrium distribution, a fact, which a rigorous non-equilibrium theory has to derive by itself.

In 1827 the English botanist Robert Brown noticed an intensive and continuous motion of small particles immersed in a rested fluid. In his honor this chaotic motion had been named Brownian. In the beginning of the 20$^{th}$ century Einstein, Smoluchowski and Langevin discovered the origin of the strange behavior of Brownian particles. Their physical theory explains the phe-nomenon as a result of interaction between the Brownian particle and continuously moving fluid particles. Due to the chaotic nature of this interaction in respect to magnitude, direction and duration, the behavior of a Brownian particle is quite irregular and can be described only in the frames of a statistical approach. The development of the theory of Brownian motion initiates a rise of many new statistical methods, which can be divided generally into two groups: probabilistic and stochastic ones.

**Probabilistic methods**

The basic feature of probabilistic methods is a replacement of the complicate mechanical description of the dynamics of interaction between the Brownian and fluid particles by some simple probabilistic assumptions about the random motion of the Brownian particle. As a result, simple linear differential equations for the evolution of the main statistical characteristics of the Brownian motion are obtained. The founder of the probabilistic methods for description of non-equilibrium processes in the Nature is Albert Einstein. In his classical paper [15] concerning the theory of Brownian motion Einstein had introduced probabilistic statistical methods for modeling of non-equilibrium processes and in this way made a start with lay the beginning of the development and application of the methods of this group. His theory of the Brownian motion is based on two intuitively clear assumptions about the motion of the Brownian particles. First, every Brownian particle does not interact with others. Second, the displacements of one and the same Brownian particle in two separate sufficiently large time intervals are statistically independent. On the basis of these two assumptions Einstein had derived numerous results becoming now classical in the theory of Brownian motion: the evolution of the Brownian particles density obeys the well-known diffusion equation, the probability density for finding a Brownian particle in a given place at a given time is Gaussian and its dispersion is proportional to time multiplied by the diffusion coefficient $D$ of the Brownian particles in the fluid, etc.

After the classical works of Einstein and Smoluchowski [16] on the Brownian motion theory lots of powerful statistical methods for solving many complicated problems have arisen. Nowadays, the most used one is the theory of Markov random processes. The main assumption of this theory is the possibility the processes in the Nature to be described via a useful idealization called stochastic processes without memory. The Markov random processes are second in respect to complexity after the random processes with independent realizations at different time moments. A characteristic feature of every Markov process is that the whole information about its contemporary evolution is contained only by the state at the last previous moment, which represents a more rigorous formulation of the Einstein assumption concerning the motion a Brownian particle. In the large variety of Markov processes there is a class of them with wide application in chemistry and physics, known as diffusive processes. The theory of Markov diffusive processes states that the time evolution of the probability density $\rho(\mathbf{x},t)$ of a random process obeys the following equation

$$\partial_t \rho = \partial_\mathbf{x} \cdot \{\mathbf{A}(\mathbf{x},t)\rho + \partial_\mathbf{x} \cdot [\mathbb{D}(\mathbf{x},t)\rho]\} \tag{1}$$

where $\mathbf{A}(\mathbf{x},t)$ and $\mathbb{D}(\mathbf{x},t)$ are the drift velocity and diffusion tensor, respectively. Equation (1), known in physics as the Fokker-Planck equation, is the main tool for modelling kinetic problems in modern science [9, 10, 14]. It is applied either in mathematics and physics or in chemistry, biology, radiotechnics, etc.

The application of the theory of Markov diffusion processes to the Brownian motion provides numerous important results. For instance, the behaviour of a free Brownian particle in its coordinate space is well described by the Wiener diffusion process. Its main characteristics are $\mathbf{A} = 0$ and $\mathbb{D} = D\mathbb{I}$ ( $\mathbb{I}$ is the unit tensor) and Eq. (1) reduces to

$$\partial_t \rho = D\partial_\mathbf{r}^2 \rho \tag{2}$$

As is seen, Eq. (2) is the classical diffusion equation, a result obtained by Einstein [15] in 1905. Usually, the application of Eq. (1) to the Brownian motion is related to a priory postulate that the equilibrium solution is the well-known Maxwell-Boltzmann distribution. This leads to a relationship between $\mathbf{A}$ and $\mathbb{D}$ called fluctuation-dissipation theorem (FDT) [17]. The diffusive process describing the evolution of a free Brownian particle in the velocity space is called the Ornstein-Uhlenbeck process [18]. Its characteristics are $\mathbf{A} = b\mathbf{v}$ and $\mathbb{D} = k_B T b \mathbb{I}/m$, where $m$ and $b$ are the mass and specific friction coefficient of the Brownian particle, and $T$ is temperature. FDT is already applied to these expressions. The Fokker-Planck equation for the Ornstein-Uhlenbeck process above reads

$$\partial_t \rho = b\partial_\mathbf{v} \cdot (\mathbf{v}\rho + k_B T \partial_\mathbf{v} \rho / m) \tag{3}$$

The equilibrium solution of Eq. (3) is the well-known Maxwell distribution.

The theory of diffusion processes provides also equations for a Brownian particle moving under the action of an external potential $U$. In this case the probability density evolution in the coordinate space obeys the following Fokker-Planck equation

$$\partial_t \rho = \partial_\mathbf{r} \cdot (\rho \partial_\mathbf{r} U + k_B T \partial_\mathbf{r} \rho) / mb \tag{4}$$

known after the name of Smoluchowski. Since FDT is already applied to Eq. (4) its equilibrium solution is the Boltzmann distribution. The juxtaposition of Eq. (4) in the case of absence of the external potential with Eq. (2) provides the Einstein relation between the diffusion and friction coefficients

$$D = k_B T / mb \tag{5}$$

Finally, the evolution of the probability density $\rho(\mathbf{v},\mathbf{r},t)$ in the Brownian particle phase space can be described by the following Fokker-Planck equation

$$\partial_t \rho + \mathbf{v} \cdot \partial_\mathbf{r} \rho - \partial_\mathbf{r} U \cdot \partial_\mathbf{v} \rho / m = b \partial_\mathbf{v} \cdot (\mathbf{v} \rho + k_B T \partial_\mathbf{v} \rho / m) \tag{6}$$

which was derived by Kramers [19]. The equilibrium density provided by Eq. (6) is the Maxwell-Boltzmann distribution, a result implied in FDT. Starting from the Kramers equation with $U = 0$ one can derive Eq. (3) by integrating only along the Brownian particle coordinate. In the literature [9, 20] there are methods proposed for derivation of the equation governing the probability density evolution in the coordinate space from Eq. (6). The result is a telegraph-like equation

$$\partial_t^2 \rho + b \partial_t \rho = \partial_\mathbf{r} \cdot (\rho \partial_\mathbf{r} U + k_B T \partial_\mathbf{r} \rho) / m \tag{7}$$

which indicates a non-Markov behaviour. Similar equation is used in the description of the turbulent diffusion [21]. In the case of adiabatic exclusion of the quick variables [22, 23] Eq. (7) reduces to the Smoluchowski equation (4).

As is seen, the probabilistic methods provide many good results in which, due to some general assumptions, the main features of the phenomenon reflect. For this reason any theory of Brownian motion has to reproduce the results obtained above. It has to be noted that the derivation of these equations can be achieved by methods based on statistical considerations [3, 19] not included in the theory of Markov diffusion processes. However, the probabilistic methods have some shortcomings: they are based on debatable presumptions about the character of the Brownian motion, which could be valid only in limited cases; there are lots of unknown constants in the kinetic equations obtained by probabilistic methods, which physical meaning and evaluation are out of this theory; the probabilistic methods are good for the most general description of Brownian motion but they cannot say anything about the detail processes leading to existence of the phenomenon, etc. These facts impose introduction of phenomenological constants and using FDT.

## Stochastic methods

The main feature of the second group of methods, the stochastic ones, is introduction of a stochastic approximation of the Brownian dynamics. In the frames of this group a set of stochastic equations is employed to obtain statistical properties of the Brownian motion or to derive equations for the probability density evolution analogous to the Fokker-Planck ones. The founder of the stochastic methods is Paul Langevin. In his classical work [24] he had introduced the first stochastic differential equation describing the dynamics of Brownian motion. Langevin divided the interaction between the Brownian and fluid particles into two parts: a resistance force, which in the frames of the hydrodynamics is given by the Stokes formula, and a fluctuation force with zero mean value, which is due to the permanent collisions between the Brownian and fluid particles. As a result of this separation and the laws of classical mechanics one can write the following equation

$$m\ddot{\mathbf{R}} + mb\dot{\mathbf{R}} + \partial_{\mathbf{R}} U = \mathbf{f} \tag{8}$$

describing the dynamics of a Brownian particle. Here $\mathbf{R}(t)$ is the coordinate of the Brownian particle. Equation (8) is the famous Langevin equation and every solution of which represents a separate realization of the random Brownian trajectories. Using relatively simple arguments about the statistical properties of the fluctuation force $\mathbf{f}$ one can obtain measurable average results for the Brownian motion. Langevin assumed that the fluctuation force and the coordinate of the Brownian particle are statistically independent random quantities and that the mean kinetic energy of the Brownian particle is equal to the average thermal energy $k_B T$. On the basis of these two assumptions and Eq. (8) he obtained an expression for the dispersion of free Brownian particles in the coordinate space

$$<\mathbf{R}^2> = 6k_B T\{t - [1 - \exp(-bt)]/b\}/mb$$

This result is more accurate than that obtained by Einstein. For times larger than the relaxation time $b^{-1}$ both expression coincide and their comparison leads to the relation (5) between the diffusion and friction coefficients. In the opposite case the Brownian particle root mean square displacement is equal to the product of the thermal velocity and time, as expected. In contrast to the previous considerations the Langevin approach provides expression for the friction constant even if a macroscopic one.

From the Langevin equation (8) one can practically derive all statistical characteristics of the motion of a Brownian particle if the statistical properties of the fluctuation force are known. The basic model for $\mathbf{f}$ is the well-known random process called white noise, which is delta correlated Gaussian process with zero mean value [25, 26]. Attaching it to Eq. (8) the theory of stochastic differential equations has arisen which is of equal worth with the Fokker-Planck equation method. There are many approaches to derive the latter from stochastic differential equations [25-27]. Nowadays, the Langevin ideas are developed and enriched by introducing some contemporary methods of non-equilibrium statistical mechanics. In the literature [17, 28, 29] an exact stochastic equation is derived which describes the Brownian particle behaviour

$$m\ddot{\mathbf{R}} + \int_0^t \mathbb{G}(t-s) \cdot \dot{\mathbf{R}}(s)ds + \partial_{\mathbf{R}}U = \mathbf{f} \qquad (9)$$

This equation is known as generalized Langevin equation (GLE). It is derived from the many particles classical mechanics using FDT only. For this reason the following statistical properties

$$<\mathbf{f}>=0 \qquad <\mathbf{f}(t)\mathbf{f}(s)>=k_B T \mathbb{G}(t-s) \qquad <\mathbf{f}(t)\dot{\mathbf{R}}(0)>=0 \qquad (10)$$

are compulsory for any stochastic theory of the Brownian motion.

The main difficulty in the application of GLE (9) is the definition of the memory function $\mathbb{G}$, which accounts for the influence of previous states on the present behaviour of the Brownian particle. There are many models for $\mathbb{G}$. Forster [29] has shown that in the case of a Brownian particle much heavier than the fluid particles the memory kernel can be well approximated by a Dirac delta-function, $\mathbb{G}(\tau) = 2mb\delta(\tau)\mathbb{I}$. Hence, in this case GLE (9) reduces to Eq. (8). In general, GLE describes non-Markov stochastic processes if the memory kernel differs from a delta-function. Some authors [13, 30] have proposed an exponentially decaying model for the memory function. For a brief review of other $\mathbb{G}$-models one can see Refs. [31, 32].

### Correspondence between the methods

As is seen, the stochastic methods describe better the Brownian motion but their application still needs FDT and phenomenological models. The work with stochastic differential equations is very helpful but not easy and acquires lots of skill and erudition. As mentioned before there are ways [25-27] starting from the stochastic equation (8) to obtain the corresponding Fokker-Planck equation (1) the use of which is much easier and transparent. There are also some methods [12, 13, 30, 33] to derive a generalized Fokker-Planck equation from GLE. Forster [29] has discussed the transition of the latter to Eq. (1) in the limit when the memory time is negligible compared to the relaxation time of the Brownian particle. Hereafter, we propose a new approach [34] for derivation of equations governing the probability density evolution. This approach is based on a special operator constructed by stochastic power expansion.

The probability density in the Brownian particle phase space $\{\mathbf{r},\mathbf{v}\}$ is given by (see for instance Ref. [10]) $\rho(\mathbf{r},\mathbf{v},t) \equiv <\delta(\mathbf{r}-\mathbf{R})\delta(\mathbf{v}-\dot{\mathbf{R}})>$, where the brackets $<\cdot>$ indicate statistical average over $\mathbf{R}$-realizations. Another useful statistical quantity is the characteristic function

$$\chi(\mathbf{q},\mathbf{u},t) \equiv <\exp(i\mathbf{q}\cdot\mathbf{R}+i\mathbf{u}\cdot\dot{\mathbf{R}})> \qquad (11)$$

The relation between $\chi$ and the probability density $\rho$ is a standard Fourier transformation and the use of one of these quantities is a matter of convenience. In the following derivation of the equation of the probability density evolution in the phase space one needs always to provide expressions for the following average product $<\mathbf{X}\delta(\mathbf{r}-\mathbf{R})\delta(\mathbf{v}-\dot{\mathbf{R}})>$, where $\mathbf{X}$ is an arbitrary random quantity. The Fourier image of this product can be presented in the form

$$<\mathbf{X}\exp(i\mathbf{q}\cdot\mathbf{R}+i\mathbf{u}\cdot\dot{\mathbf{R}})>=<\sum\frac{1}{j!k!}<\mathbf{X}[i\mathbf{q}\cdot(\mathbf{R}-\mathbf{R}')]^{j}[i\mathbf{u}\cdot(\dot{\mathbf{R}}-\dot{\mathbf{R}}')]^{k}>\exp(i\mathbf{q}\cdot\mathbf{R}'+i\mathbf{u}\cdot\dot{\mathbf{R}}')>'$$

In fact the right hand side of this equation is a power expansion of the trajectory $\mathbf{R}$ around another realization $\mathbf{R}'$ of the same random process, which is the essence of the proposed method. Taking inverse Fourier transformation of this equality results in

$$<\mathbf{X}\delta(\mathbf{r}-\mathbf{R})\delta(\mathbf{v}-\dot{\mathbf{R}})>=\hat{P}(\mathbf{X})\rho(\mathbf{r},\mathbf{v},t) \qquad (12)$$

where the operator $\hat{P}(\mathbf{X})$ is defined via the relation

$$\hat{P}(\mathbf{X})\rho=\sum\frac{1}{j!k!}(\partial_{\mathbf{r}}\cdot)^{j}(\partial_{\mathbf{v}}\cdot)^{k}[<(\mathbf{r}-\mathbf{R})^{j}(\mathbf{v}-\dot{\mathbf{R}})^{k}\mathbf{X}>\rho]$$

This operator possesses an important property: if the quantity $\mathbf{X}$ satisfies the following correlation properties

$$<\mathbf{R}^{j}\dot{\mathbf{R}}^{k}\mathbf{X}>=\frac{j!}{J!(j-J)!}\frac{k!}{K!(k-K)!}<\mathbf{R}^{J}<\mathbf{R}^{j-J}\dot{\mathbf{R}}^{k-K}>\dot{\mathbf{R}}^{K}\mathbf{X}>$$

for $j\geq J$ and $k\geq K$, and $<\mathbf{R}^{j}\dot{\mathbf{R}}^{k}\mathbf{X}>=0$ for $j<J$ and $k<K$, then its corresponding operator reduces to

$$\hat{P}(\mathbf{X})=\frac{(-1)^{J+K}}{J!K!}<\mathbf{X}(\mathbf{R}\cdot\partial_{\mathbf{r}})^{J}(\dot{\mathbf{R}}\cdot\partial_{\mathbf{v}})^{K}> \qquad (13)$$

Analogically one can introduce operators in the coordinate and velocity space, respectively. For instance, the coordinate operator reads

$$\hat{R}(\mathbf{X})\rho(\mathbf{r},t)=\sum\frac{1}{j!}(\partial_{\mathbf{r}}\cdot)^{j}[<(\mathbf{r}-\mathbf{R})^{j}\mathbf{X}>\rho(\mathbf{r},t)]=\int_{-\infty}^{\infty}\hat{P}(\mathbf{X})\rho(\mathbf{r},\mathbf{v},t)d\mathbf{v}$$

and as is seen it is simply related to the operator $\hat{P}(\mathbf{X})$.

The first demonstration of the proposed method is a derivation of the equation describing the evolution of the probability density in the phase space of the Brownian particle. Taking a time derivative of the characteristic function (11) and using Eq. (8) yields the following equation

$$\partial_t \chi = <[i\mathbf{q}\cdot\dot{\mathbf{R}} - i\mathbf{u}\cdot(mb\dot{\mathbf{R}} + \partial_\mathbf{R} U - \mathbf{f})/m]\exp(i\mathbf{q}\cdot\mathbf{R} + i\mathbf{u}\cdot\dot{\mathbf{R}})>$$

Applying the inverse Fourier transformation to this equation and using definition (12) one derives the evolution equation for the probability density $\rho(\mathbf{r},\mathbf{v},t)$ in the phase space

$$\partial_t \rho + \mathbf{v}\cdot\partial_\mathbf{r}\rho - \partial_\mathbf{r} U\cdot\partial_\mathbf{v}\rho/m = b\partial_\mathbf{v}\cdot[\mathbf{v}\rho - \hat{P}(\mathbf{f})\rho/mb] \tag{14}$$

Equation (14) is exact but requires knowledge about the statistical properties of the Langevin force. First, one can suppose [24] that the force $\mathbf{f}$ is not correlated with the position of the Brownian particle, i.e. $J = 0$. The second presumption is more intuitive and says that the fluctuation power $\mathbf{f}\cdot\dot{\mathbf{R}}$ is not correlated to the velocity of the Brownian particle. This assumption, combined with the relation $<\mathbf{f}>=0$, leads to accept the value $K=1$. Finally, if the average energy of a Brownian particle is stationary it follows from Eq. (8) $<\mathbf{f}\dot{\mathbf{R}}>= mb<\dot{\mathbf{R}}\dot{\mathbf{R}}>= bk_B T\mathbb{I}$. Hence, the operator $\hat{P}(\mathbf{f})$ from Eq. (13) acquires the form

$$\hat{P}(\mathbf{f}) = -bk_B T\partial_\mathbf{v} \tag{15}$$

The combination of Eqs. (14) and (15) is the Kramers equation (6).

The derivation of Eq. (3) by the present method is obvious. It is more interesting how the evolution of the probability density in the coordinate space could be described. The corresponding characteristic function is $\chi(\mathbf{q},0,t)$. Taking a first derivative in respect to time of this function and inverting the Fourier image yields

$$\partial_t \rho = -\partial_\mathbf{r}\cdot[\hat{R}(\dot{\mathbf{R}})\rho]$$

This equation is the well-known continuity equation. By appropriate modelling of the operator $\hat{R}(\dot{\mathbf{R}}) = -D\partial_\mathbf{r}$ one can derive the diffusion equation (2). However, more detailed considerations require the use of the second time-derivative of the characteristic function and the stochastic differential equation (8). The result is

$$\partial_t^2 \chi(\mathbf{q},0,t) = i\mathbf{q}\cdot <(i\mathbf{q}\cdot\dot{\mathbf{R}}\dot{\mathbf{R}} - b\dot{\mathbf{R}} - \partial_{\mathbf{R}}U/m + \mathbf{f}/m)\exp(i\mathbf{q}\cdot\mathbf{R})>$$

and applying the inverse Fourier transformation one can derive in a general form the differential equation governing the evolution of the probability density in the coordinate space

$$\partial_t^2\rho + b\partial_t\rho = \partial_{\mathbf{r}}\cdot\{\rho\partial_{\mathbf{r}}U/m + \partial_{\mathbf{r}}\cdot[\hat{R}(\dot{\mathbf{R}}\dot{\mathbf{R}})\rho] - \hat{R}(\mathbf{f})\rho/m\} \qquad (16)$$

This equation requires some modelling of the included operators. As a consequence of the statistical properties of the Langevin force it follows from Eq. (15) that $\hat{R}(\mathbf{f})=0$. Regarding the operator $\hat{R}(\dot{\mathbf{R}}\dot{\mathbf{R}})$ it is reasonable to accept that the kinetic energy is not correlated to the trajectory of the Brownian particle, i.e. $J=0$. Therefore, this operator acquires the simple form $\hat{R}(\dot{\mathbf{R}}\dot{\mathbf{R}}) = <\dot{\mathbf{R}}\dot{\mathbf{R}}> = k_B T \mathbb{I}/m$ and Eq. (16) reduces to Eq. (7). Considering now the case of a free Brownian particle ($U=0$) Eq. (7) can be rewritten as

$$\partial_t\rho = D\Box\rho$$

where $\Box$ is a d'Alembert operator with specific velocity being the thermal velocity of the Brownian particle. The presence of the second time-derivative here is an indication for a limit in the speed of the process. In the diffusion the thermal velocity is the highest possible one for the particle transport and plays a restrictive role for the process like the velocity of light is the maximal one for energy transfer in the world.

## Appendix

This appendix is added to the printed paper to show via the central limit theorem that the white noise is compulsory Gaussian due to its continuity, stationarity and constant spectral density. In the recent years the interest to non-Gaussian white noises has increased as a tool for description of more complex stochastic systems. The present appendix proves that the white noise is only Gaussian and, hence, non-Gaussian white noises are oxymora.

The white noise is a stationary stochastic process $f(t)$, which is zero centred $<f>=0$ and delta-correlated $<f(t_1)f(t_2)>=\delta(t_1-t_2)$. Its probability distribution is not specified. A typical example for a white noise is the classical Langevin force [24]. One can present $f(t)$ by its Fourier image $g(\nu)$ via the standard integral relation

$$f(t) = \int_{-\infty}^{\infty} g(\nu)\exp(2\pi i\nu t)d\nu \qquad (A1)$$

Using the inverse Fourier transformation of Eq. (A1) and the white noise mean value and auto-correlation function, it follows that $<g>=0$ and $<g(\nu_1)\overline{g}(\nu_2)>=\delta(\nu_1-\nu_2)$ [9], i.e. the Fourier image is also a white noise. Thus, according to these expressions and Eq. (A1) the white noise $f$ represents an integral sum of infinite number uncorrelated Fourier components possessing zero mean value and the same dispersion. Note that the Fourier images are not correlated for any stationary stochastic process but only for the white noise they possess equal dispersions [9]. Hence, the necessary conditions for application of the central limit theorem are fulfilled and the conclusion is that the white noise is compulsory Gaussian. The main reason for that is the equal weight of the Fourier components in Eq. (A1), which is the reason for calling the noise $f(t)$ white in analogy to the white light.